\documentclass{edm_article}
\usepackage{amsmath,amsfonts}
\usepackage{algorithmic}
\usepackage{algorithm}
\usepackage{array}
\usepackage[caption=false,font=normalsize,labelfont=sf,textfont=sf]{subfig}
\usepackage{textcomp}
\usepackage{stfloats}
\usepackage{url}
\usepackage{verbatim}
\usepackage{graphicx}
\usepackage{booktabs}
\usepackage{hyperref}
\usepackage{cite}
\usepackage{array, makecell}
\usepackage{mathtools}
\usepackage[x11names,table]{xcolor}
\usepackage{tabularx}
\usepackage[protrusion=true,expansion=true]{microtype}

\definecolor{algoCol}{HTML}{FFCCCC}   
\definecolor{compCol}{HTML}{CCCCFF}   
\definecolor{shakeCol}{HTML}{CCFFCC}  
\definecolor{mlCol}{HTML}{FFFFCC}

\begin{document}

\title{Spatiotemporal Link Formation Prediction in Social Learning Networks Using Graph Neural Networks}


%
%
%
%

\numberofauthors{4} 
%
\author{
%
%
\alignauthor
Ali Mohammadiasl\\
       \affaddr{UC San Diego}\\
       \email{amohammadiasl@ucsd.edu}
\alignauthor
Bita Akram\\
       \affaddr{NC State University}\\
       \email{bakram@ncsu.edu}
\and  
\alignauthor 
Seyyedali Hosseinalipour\\
       \affaddr{University at Buffalo--SUNY}\\
       \email{alipour@buffalo.edu}
\alignauthor Rajeev Sahay\\
       \affaddr{UC San Diego}\\
       \email{r2sahay@ucsd.edu}
}

\maketitle

\begin{abstract}

Social learning networks (SLNs) are graphical representations that capture student interactions within educational settings (e.g., a classroom), with nodes representing students and edges denoting interactions. Accurately predicting future interactions in these networks (i.e., link prediction) is crucial for enabling effective collaborative learning, supporting timely instructional interventions, and informing the design of effective group-based learning activities. However, traditional link prediction approaches are typically tuned to general online social networks (OSNs), often overlooking the complex, non-Euclidean, and dynamically evolving structure of SLNs, thus limiting their effectiveness in educational settings. In this work, we propose a graph neural network (GNN) framework that jointly considers the temporal evolution within classrooms and spatial aggregation across classrooms to perform link prediction in SLNs. Specifically, we analyze link prediction performance of GNNs over the SLNs of four distinct classrooms across their (i) temporal evolutions (varying time instances), (ii) spatial aggregations (joint SLN analysis), and (iii) varying spatial aggregations at varying temporal evolutions throughout the course. Our results indicate statistically significant performance improvements in the prediction of future links as the courses progress temporally. Aggregating SLNs from multiple classrooms generally enhances model performance as well, especially in sparser datasets. Moreover, we find that jointly leveraging both the temporal evolution and spatial aggregation of SLNs significantly outperforms conventional baseline approaches that analyze classrooms in isolation. Our findings demonstrate the efficacy of educationally meaningful link predictions, with direct implications for early-course decision-making and scalable learning analytics in and across classroom settings.

\end{abstract}

\keywords{Graph Neural Networks, Link Prediction, Social Learning Networks, Social Network Analysis.} 

\section{Introduction} \label{sec:Introduction}



Collaborative learning is a major pillar of modern education, shaping student engagement, persistence, and academic success. Consequently, instructors increasingly incorporate structured peer interactions, such as peer instruction and discussion forums, to enhance learning outcomes \cite{sna_toe,colab_learning}. Yet, in practice, collaborative behaviors do not emerge uniformly across students or over time: some students rapidly become socially central, while others remain weakly connected or isolated, particularly in the early stages of a course. These uneven and evolving interaction patterns complicate instructors’ ability to anticipate collaboration dynamics and intervene effectively when support is most needed.

In order to model and analyze such interaction dynamics, social learning networks (SLNs) have emerged as a powerful abstraction. SLNs represent classroom interactions as graphs, where \textit{nodes} correspond to students and \textit{edges} denote interactions arising from discussions, question–answer exchanges, or collaborative activities \cite{SLN_benefits, SocialNetworksAsLearningEnvironment}. This abstraction enables a principled investigation of network inference problems in educational graphs, with \textit{link prediction} (i.e., forecasting which student interactions are likely to form in the future) serving as one of the fundamental problems in this setting.


 \vspace{-1mm}
\subsection{Challenges and Motivations}\label{sec:Hypoth}
Within SLNs, \textit{link prediction} has emerged as a valuable tool for forecasting future collaborations, identifying at-risk students, and informing instructional interventions. Furthermore, accurate link prediction can reveal relationships between students' behavioral characteristics and their tendencies toward class involvement, which is often linked to their learning performance \cite{9162494}. However, prior work on link prediction in educational settings has largely focused on analyzing each classroom independently, treating the SLN associated with a single course offering as an isolated graph \cite{InfocomeDataset,OG_Paper}. While this approach is intuitive, it implicitly assumes that meaningful structure must be learned anew for every classroom, ignoring common patterns of social and pedagogical dynamics that recur across courses, instructors, and academic terms. In addition, this approach fails to capture the temporal processes through which collaborations form early in the course when few observed links have been formed (i.e., the cold-start problem), thus preventing accurate link prediction precisely when early intervention would be most beneficial. Moreover, the inherent complexity of SLNs, including their evolving topology and irregular non-Euclidean structure, renders traditional data-driven models \cite{ieee_dynamic_patterns} \cite{acm_temporal_patterns}, such as convolutional neural networks (CNNs), ill-equipped for the task of link prediction. 

Graph Neural Networks (GNNs), on the other hand, offer a powerful alternative to traditional CNNs when it comes to predicting links in SLNs. Unlike CNNs, which are primarily designed for grid-like data such as images, GNNs are tailored for irregular graph structures. This makes them particularly well-suited for SLNs, where nodes (students) and edges (interactions) form complex non-Euclidean structures \cite{acm_graphrec}. Despite their potential, the application of GNNs in SLNs remains in its early stages, with only a handful of studies exploring their potential and limitations~\cite{OG_Paper}. 

To address this gap in the literature, \textit{our work leverages GNNs to unlock two critical domains in link prediction for SLNs: the `temporal' domain and the `spatial' domain.} Specifically, in the temporal domain, we examine the evolution of SLN graphs over time, while in the spatial domain, we investigate the effects of combining SLN graphs from different classroom contexts to improve link prediction performance equitably across classrooms.

     \vspace{-1mm}
\subsubsection{Exploring the Temporal Domain} SLNs are inherently dynamic: the topology of these networks evolves as student interactions change throughout the course. This temporal evolution can significantly affect GNNs' future link prediction performance in SLNs. By analyzing how SLN structures shift at different stages of a course, we can uncover deeper insights into student interactions and the conditions under which link prediction performs best.
We subsequently present and investigate, for the first time in the literature, the following hypothesis:

 \vspace{-1mm}
\underline{\textbf{\textit{Hypothesis 1:}}} \textit{Do SLNs' temporal structures have a significant impact on the prediction capability of GNNs trained on them? If yes/no, how do link prediction performance metrics vary during the temporal evolution of SLNs?}

 \vspace{-1mm}
\subsubsection{Investigating the Spatial Domain}
Our study on the spatial domain of SLNs introduces the combination of SLN graphs across different classrooms. At first glance, merging SLNs might seem to improve model performance by providing more data. However, our intuition suggests that the unique topological structures inherent in each SLN, stemming from varied interaction patterns among peers in different classrooms, can lead to unexpected outcomes.
Subsequently, our formally proposed hypothesis that we investigate for the first time in the literature is as follows:

 \vspace{-1mm}
\underline{\textbf{\textit{Hypothesis 2:}}} \textit{Does combining SLNs together lead to better performance of the GNN models trained on them? If yes/no, how do link prediction performance metrics vary upon combining SLNs?}

 \vspace{-1mm}
\subsubsection{Spatiotemporal Analysis}

A careful observation of the interrelations between the temporal and spatial domains reveals non-trivial interactions. For instance, our intuition may suggest that combining SLNs at earlier stages of a course might be more beneficial, as networks are still forming and less defined. However, as SLNs develop their distinct patterns later in the course, merging them might not yield the same performance improvements. To encapsulate this nuanced view, we propose and investigate the following hypothesis:

 \vspace{-1mm}
\underline{\textbf{\textit{Hypothesis 3:}}} \textit{Does combining SLNs together at different time stamps during a course lead to varying performance of the GNN models trained on them? If yes/no, how do link prediction performance metrics vary during the combination of SLNs at different time stamps of a course?}

\subsection{Overview and Summary of Contributions}

Formally, our contributions can be summarized as follows:
 \vspace{-3.5mm}
\begin{itemize}
    \item We pioneer an investigation into the temporal evolution and spatial aggregation of SLNs by analyzing how link formation and interaction patterns change throughout a class and how classrooms can benefit one another through sharing their data. Our time-evolving analysis over multiple classroom SLNs provides novel insights into these dynamics.
    \item We develop a GNN-based framework designed to tackle the unique challenges presented by sparse, imbalanced, and temporally evolving SLN datasets. This framework is optimized to capture the complex topological features inherent in SLNs. We then use this framework to validate our \textbf{\textit{Hypothesis 1}}.
    \item We demonstrate that combining data from multiple SLNs, spanning different classroom contexts, can enhance link prediction performance under certain conditions. This contribution is focused on validating our \textbf{\textit{Hypothesis 2}}. 
    \item By examining the interplay between the temporal evolution and spatial data aggregation of SLNs, we uncover non-trivial interactions that influence overall link prediction accuracy. This integrated analysis addresses our \textbf{\textit{Hypothesis 3}}, where our findings offer a refined understanding of how temporal and spatial factors collectively inform link prediction strategies.
\end{itemize}



\section{Related Work} \label{related_work}

Link prediction in general computational graphs (i.e., non-SLN graphs such as Online Social Networks (OSNs)) has evolved through various methodologies over time. Early methods relied on deriving features from the graph topology \cite{SLN_LinkPrediction, Pairwise} to infer future trends \cite{WL_NM, Community-enhanced}. While effective in some cases, these heuristic-based methods often fail to capture the inherent complexities of dynamic, rapidly-evolving graphs. To address these limitations, more rigorous link prediction approaches were developed, such as matrix factorization and probabilistic models, which aim to learn low-dimensional representations of nodes \cite{Koren_MatrixFactorization_2009}, \cite{IEEE_ProbabilisticModels}, \cite{ACM_ProbModels}, \cite{zhu2016maxmarginnonparametriclatentfeature}. Although these models yielded improved prediction accuracy, later methods further advanced performance by offering greater scalability and adaptability to capture the complex, evolving nature of OSNs. In particular, the advent of feature-based deep learning introduced models based on artificial neural networks for link prediction, which leverage node characteristics and graph adjacency information to improve link prediction in both general OSNs \cite{MLP_models,CNN_LinkPrediction_2016,DeepLink,mohamady2024link} as well as SLNs \cite{hridi2025privacy,OG_Paper,InfocomeDataset}. While these methods achieved notable success, they typically treat graphs as static structures and lack the capability to effectively capture the temporal dynamics inherent in SLNs. Moreover, they fail to account for the intricate topological structures and co-temporal evolution of interactions across multiple classrooms, limiting their applicability in educational settings.

More recently, GNNs have emerged as powerful tools for link prediction in general graph structures due to their ability to learn both the local structure and the global topology of a graph \cite{xu2021topic,sharma2024survey,khoshraftar2024survey,sankar2021graph,li2023evaluating,chen2024social,dileo2024temporal}. GNN-based models, such as Graph Convolutional Networks (GCNs) \cite{Kipf_GCN_2017}, GraphSAGE \cite{Hamilton_GraphSAGE_2017}, and Graph Attention Networks (GATs) \cite{Velickovic_GAT_2018}, have demonstrated superior performance compared to feature-based deep learning models by aggregating neighborhood information through learned message-passing schemes among nodes. Despite these advancements, no prior work has specifically examined GNN-based link prediction approaches in dynamically evolving SLNs, thus overlooking the temporal evolution and spatial aggregation properties that are unique to SLNs. As a result, the impact of SLN dynamics over time, the integration of information across classroom contexts, and the interplay between these two dimensions remain largely unexplored. Our work aims to address this critical gap in the literature.







\section{Methodology} \label{methods}

In this section, we begin by describing the mathematical representation of SLNs as graphs (Sec. \ref{sln_graph}). Next, we outline the task of link prediction in SLNs, detailing how GNNs are used to infer future links (Sec. \ref{link_prediction}). We then introduce our joint classroom approach, detailing the  integration of SLNs to improve predictive performance in individual classrooms (Sec. \ref{joint_classroom}). Lastly, we describe the evaluation metrics used to assess the link prediction performance  (Sec. \ref{evaluation}).

\subsection{Representation of SLNs as Graphs} 
\label{sln_graph}

We represent each SLN as a graph $G = (\mathcal{V}, \mathcal{E})$, where $\mathcal{V}$ denotes the set of \textit{nodes} (i.e., students) and $\mathcal{E}$ represents the set of \textit{edges} (i.e., interactions between students). Each edge $e_{ij} \in \mathcal{E}$ signifies an interaction between two nodes/students $v_i$ and $v_j$, where $v_i,v_j \in \mathcal{V}$, such as collaborative projects, discussions, or information exchanges. Subsequently, we define the adjacency matrix  of the SLN graph as $\mathbf{A}=[A_{ij}]_{v_i,v_j\in\mathcal{V}}$, where
\begin{equation}     A_{ij} =
    \begin{cases} 
    1, & \text{if } (v_i, v_j) \in \mathcal{E} \\
    0, & \text{otherwise}.
    \end{cases} \label{graph}
\end{equation}
In words, $A_{ij} $ represents the presence ($A_{ij}=1 $) or absence ($A_{ij}=0 $) of a connection/edge between nodes $v_i $ and $v_j $.

\begin{figure}[t!]
    \includegraphics[width=0.45\textwidth]{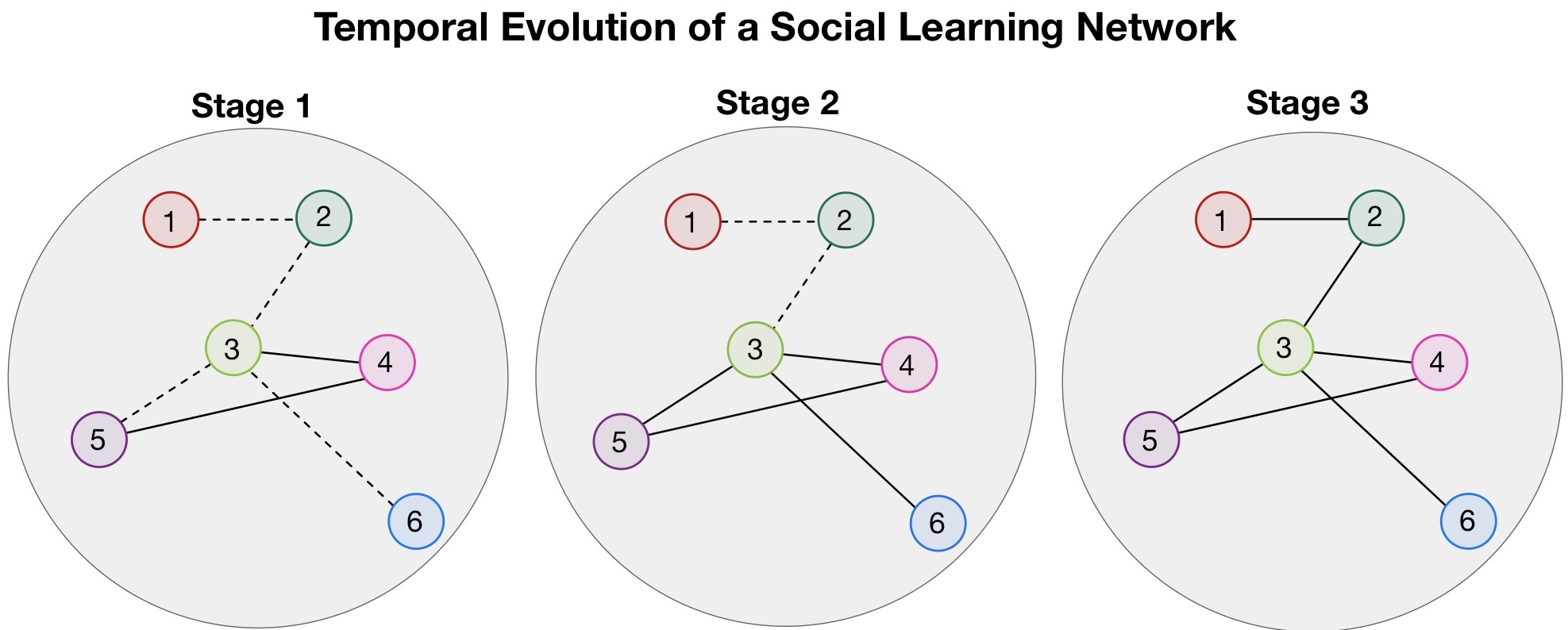}
    \caption{Visualization of SLN temporal evolution. Left to right shows the progress of the classroom from the beginning to the end. A solid line denotes an existing edge at that stage, and a dotted line denotes a non-existing edge at that stage that will become an existing one in a future stage. }
    \Description{Visualization of SLN temporal evolution. Left to right shows the progress of the classroom from the beginning to the end. A solid line denotes an existing edge at that stage, and a dotted line denotes a non-existing edge at that stage that will become an existing one in a future stage. }
    \label{fig:slnevolution}
\end{figure}
\subsection{Link Prediction in SLNs} \label{link_prediction}
The temporal evolution of an SLN is illustrated in Fig.~\ref{fig:slnevolution}, showcasing a simplified example of an SLN across three sequential stages of a course. 
Subsequently, given the SLN graph described in Sec.~\ref{sln_graph} at some point $t$ during the duration of a course/class (e.g., week 4 of the semester), our objective is to predict which links will form at a future time $t+t_0$ (e.g., week 6) between students by using GNNs. To this end, we adopt GraphSAGE \cite{Hamilton_GraphSAGE_2017}, which learns node embeddings/features by iteratively aggregating information from local neighborhoods. As new nodes are not introduced in the SLN, this inductive learning approach enables the model to dynamically adapt node embeddings as local structures (i.e., student interaction patterns) evolve over time, making it well-suited for the temporal changes of student interactions throughout the course.





\begin{figure}[t!]
    \centering
    \includegraphics[width=0.45\textwidth]{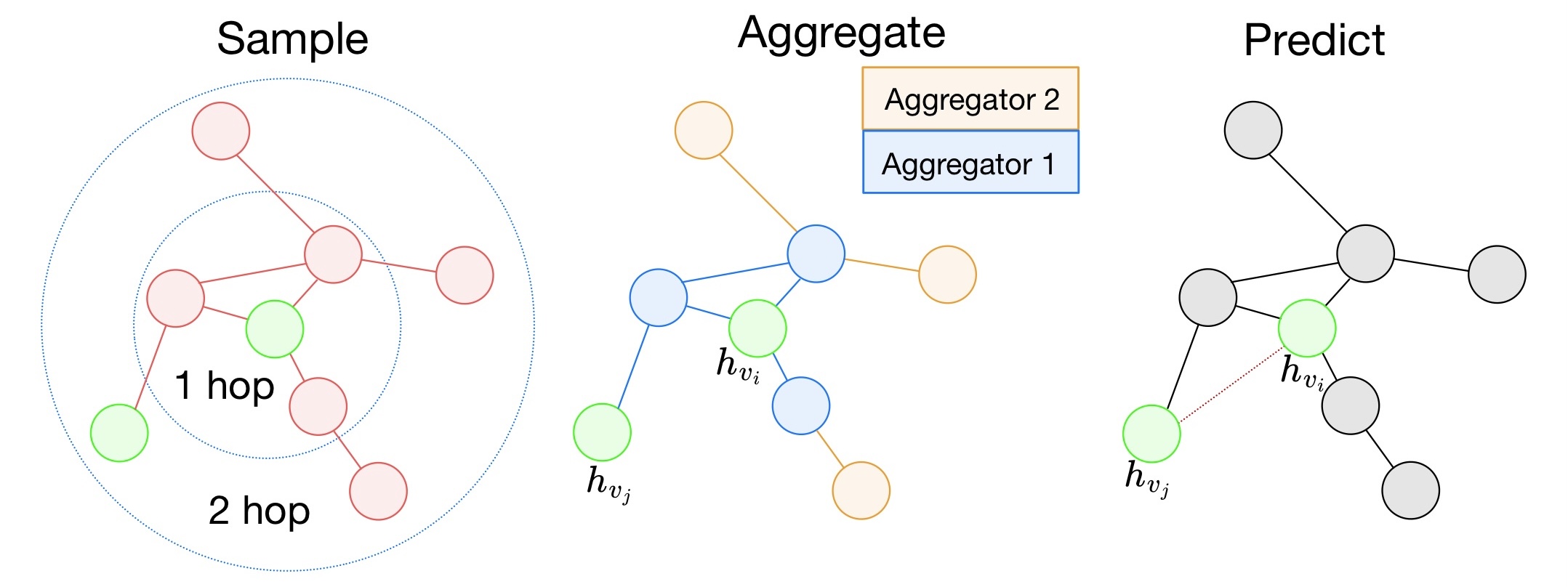}
    \caption{An overview of the GraphSAGE pipeline for link prediction. \textbf{Left:} Sampling stage, where the 1-hop and 2-hop neighborhoods of the target nodes (highlighted in green) are collected. \textbf{Middle:} Aggregation stage, where each node's embedding is updated by aggregating the features of its neighbors using multiple layers of aggregation (in our case, mean aggregator). \textbf{Right:} Prediction stage, where a dot product is computed between the final embeddings of the two target nodes to predict the existence of a link.}
    \Description{An overview of the GraphSAGE pipeline for link prediction. \textbf{Left:} Sampling stage, where the 1-hop and 2-hop neighborhoods of the target nodes (highlighted in green) are collected. \textbf{Middle:} Aggregation stage, where each node's embedding is updated by aggregating the features of its neighbors using multiple layers of aggregation (in our case, mean aggregator). \textbf{Right:} Prediction stage, where a dot product is computed between the final embeddings of the two target nodes to predict the existence of a link.}
    \label{fig:graphsage}
\end{figure}


\subsubsection{Link Predictions based on GNNs in a Nutshell} In a nutshell, to predict whether an interaction (i.e., a link) will form between two students $v_i $ and $v_j $, we compute a numerical representation (embedding) for each student using a GraphSAGE model as we will later discuss (see Sec.~\ref{sec:3.22}). These embeddings capture the structural context of each student within the SLN. The likelihood of a future interaction is then estimated by taking the dot product of the respective node embeddings (potentially after propagating through one or more hidden layer) as follows:
\begin{equation}\label{eq:finaldot}
    \hat{y}_{{v_i},{v_j}} = \sigma(h_{v_i}^\top h_{v_j}),
\end{equation}
where $\sigma(\cdot)$ is the sigmoid function, $h_{v_i}$ and $h_{v_j}$ are the learned embeddings for students $v_i$ and $v_j$, respectively ($h_{v_i},h_{v_j} \in \mathbb{R}^{d}$), and $\hat{y}_{{v_i},{v_j}} \in \mathbb{R}$. 

\subsubsection{Obtaining Node Embeddings via GNN}\label{sec:3.22} To obtain the node embeddings, we use a \textit{two-layer GraphSAGE model}, where each layer aggregates information from a student's neighbors in the SLN. This model, depicted in Fig.~\ref{fig:graphsage}, replicates a convolutional filter that is capable of  operating in a non-Euclidean space (i.e., graph topology). Intuitively, at each SAGEConv layer, GraphSAGE updates the representation of a node by combining its own features with the aggregated features of its neighbors. By the second layer, the embedding of a node reflects not only its own characteristics but also the structure and features of its broader local neighborhood, enabling the model to better infer potential future connections among nodes. 
In particular, the model begins with an input feature matrix $\mathbf{X}=[x_{v_i}]_{v_i\in\mathcal{V}} \in \mathbb{R}^{|\mathcal{V}| \times F}$, where the $i^{\text{th}}$ row (i.e.,  $ x_{v_i} \in \mathbb{R}^F$) corresponds to the feature vector of student $v_i$ consisting of $F$ features. 
Since our SLNs do not contain explicit node features (e.g., profile or grade data), each student is initialized with a unique trainable embedding vector. These initial embeddings are then refined through neighbor aggregation as follows.\footnote{Fig.~\ref{fig:graphsage} illustrates GraphSAGE assuming aggregation over all neighbors for clarity. However, in practice, GraphSAGE samples a fixed-size subset of neighbors at each layer to reduce computational overhead and memory usage.} 


In the first layer, the model updates each student’s embedding by averaging its own feature vector with those of its directly connected peers (1-hop neighbors). This step captures local interaction patterns through the following embedding:
\begin{equation}\label{eq:3}
   \hspace{-2mm} h_{v_i}^{(1)} = \zeta \Big(W_1 \cdot \text{MEAN} \left( \big\{ x_{v_j} {\mid} v_j \in \mathcal{N}(v_i) \cup \{v_i\} \big\} \right) + b_1 \Big), \hspace{-2mm}
\end{equation}
where $\zeta(\cdot)$ is the activation function, $h_{v_i}^{(1)} \in \mathbb{R}^{d_1}$ is the updated embedding for student $v_i$ after one round of aggregation, 
$\mathcal{N}(v_i)$ is the set of students directly connected to $v_i$ in the SLN (i.e., students who have interacted with $v_i$), 
$x_{v_j} \in \mathbb{R}^{F}$ is the feature vector of student $v_j$, 
$W_1$ and $b_1$ are trainable weights and bias parameters, 
and $\text{MEAN}(\cdot)$ takes the element-wise average across the set.

In the second layer, the model does the same aggregation, except this time, it uses the embedding from the first layer, allowing to implicitly capture information from 2-hop neighbors. 
In particular, the second layer of the model captures the following embedding:
\begin{equation}\label{eq:4}
   \hspace{-3mm} h_{v_i}^{(2)} {=} \zeta \left(W_2 \cdot \text{MEAN} \left( \{ h_{v_j}^{(1)} {\mid} {v_j} \in \mathcal{N}({v_i}) \cup \{{v_i}\} \} \right) + b_2 \right)\hspace{-1mm},\hspace{-3mm}
\end{equation}
where $h_{v_i}^{(2)} $ is the final embedding for student ${v_i} $ (i.e., used as $h_{v_i}$  in~\eqref{eq:finaldot}), and $W_2 $ and $b_2 $ are  trainable parameters.


\subsection{Classroom Merging and GNN Training} \label{joint_classroom}


\begin{figure}[t!]
    \includegraphics[width=0.45\textwidth]{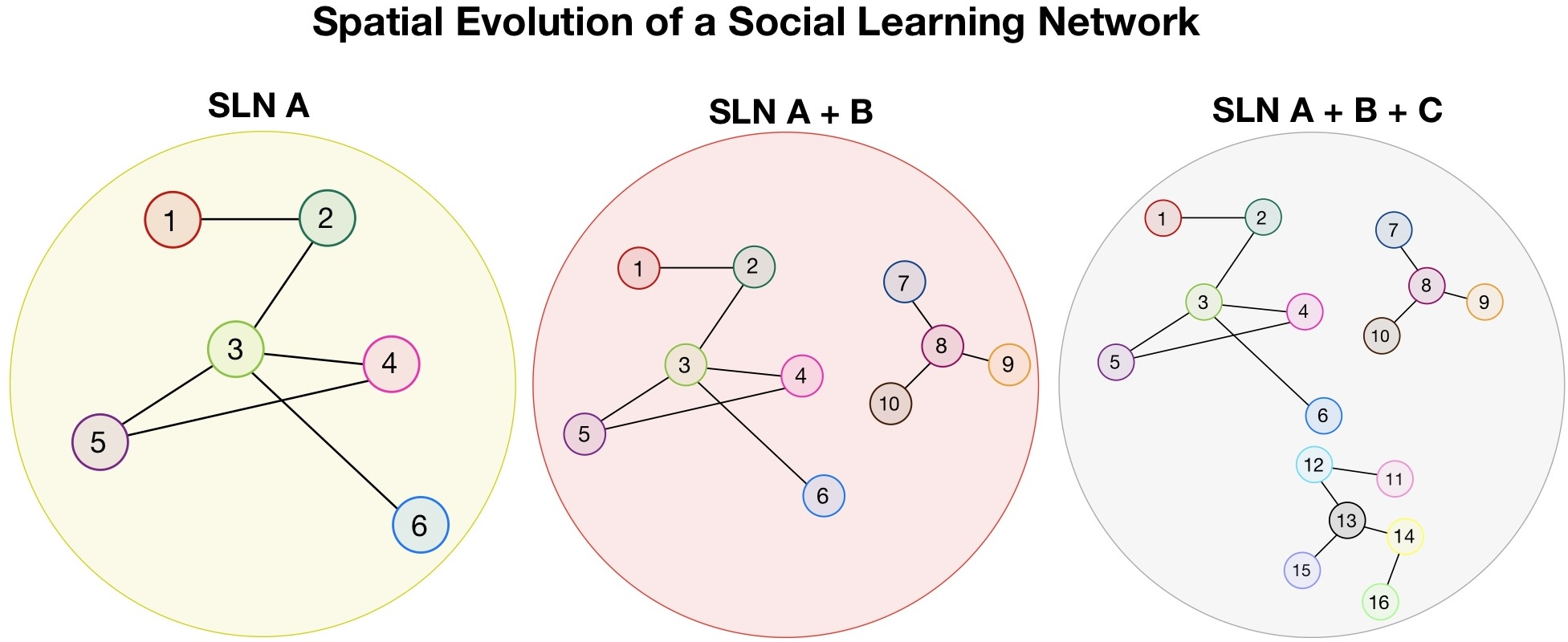}
    \caption{Visualization of spatial evolution. The left displays a single SLN A (representing a single classroom), the middle is SLN A combined with SLN B (i.e., two classrooms combined together), and the right is SLN  A and B combined with SLN C (i.e., three classrooms combined together). }
    \Description{Visualization of spatial evolution. The left displays a single SLN A (representing a single classroom), the middle is SLN A combined with SLN B (i.e., two classrooms combined together), and the right is SLN  A and B combined with SLN C (i.e., three classrooms combined together). }
    \label{fig:slnSPATIALevolution}
\end{figure}



\subsubsection{Merging Classrooms}To assess the impact of training a GNN for link prediction, we conduct experiments on individual and joint SLNs, capturing their spatial evolution. In particular, initially, we train separate GNN models for each classroom, using the same time instances of the SLN graph across the classrooms to ensure consistency across experiments. We then construct combined SLN graphs by merging classroom SLNs, systematically evaluating the combination of multiple classroom SLNs into a single joint graph. Subsequently, we train GNNs on these combined SLNs and evaluate their performance on individual classrooms to verify the impact of classroom combination on the link prediction performance of the trained models. We visualize the spatial aggregation of SLNs in Fig.~\ref{fig:slnSPATIALevolution}.

\subsubsection{GNN Training} To train GNNs over our SLNs (either on isolated or merged classrooms), we first note that there is an extreme class imbalance inherent in SLNs. This imbalance arises because \textit{SLNs are sparse graphs}: the vast majority of possible node pairs in the graph do not share an edge, meaning that most students do not interact with each other. 

To mitigate this imbalance during training, we use a weighted binary cross-entropy loss: for a predicted link score $s$ of a given node pair and the respective ground truth label $y \in \{0,1\}$ ($1$ for edge, $0$ for non-edge), the loss is defined as:
\[
\ell(s, y) =
\begin{cases}
\text{PCW} \cdot \log\big(1 + e^{-s^+}\big), & \text{if } y=1, \\
\log\big(1 + e^{s^-}\big), & \text{if } y=0.
\end{cases}
\]
The total loss across a positive and negative sample pair is:
\[
\mathcal{L} = \text{PCW} \cdot \log\big(1 + e^{-s^+}\big) + \log\big(1 + e^{s^-}\big),
\]
where $s^+$ and $s^-$ denote the scores for a positive (i.e., connected) and negative (i.e., not connected) student pair, respectively. We  found through a grid search, optimizing for the minimization of false negatives, that fixing \textit{positive class weight} $\text{PCW} = 20$ provides a stable and effective training regime across all SLNs, balancing the impact of class imbalance without over-amplifying the minority class loss.




\subsection{Evaluation Metrics} \label{evaluation}


Although accuracy is a conventional metric for classification tasks, it is not appropriate in link prediction due to the large class-imbalance, where a model with relatively high accuracy may fail to predict any edges. Instead, we use the \textit{Area Under the receiver operating characteristic Curve (AUC)}, as it provides a robust measure of predictive effectiveness in class-imbalanced settings since it evaluates a model’s ability to rank positive samples (i.e., student pairs with edges) higher than negative ones, making it a more reliable indicator of performance in sparse graphs. The AUC is computed by
\begin{equation}
    \text{AUC} = \frac{1}{|\mathcal{P}| |\mathcal{N}|} \sum_{(s^+, s^-) \in (\mathcal{P}, \mathcal{N})} \mathbb{I}(s^+ > s^-),
\end{equation}
where $\mathcal{P}$ and $\mathcal{N}$ denote the sets of predicted scores for positive and negative samples, respectively, and $\mathbb{I}(\cdot)$ is an indicator function that returns 1 if the positive score $s^+$ is greater than the negative score $s^-$, and 0 otherwise. In our context, 
the model assigns a score to each link (i.e., a student pair), where a higher score indicates a higher likelihood that the link should exist and a threshold score of $0.5$ is used to determine the prediction of a link. 





\section{Results and Discussion} \label{results}

Henceforth, we discuss our empirical setup (Sec. \ref{emp_setup}) followed by analyzing our results in temporal (Sec. \ref{tmp}), spatial (Sec. \ref{spatial}), and spatiotemporal settings (Sec. \ref{joint}). The code used in this work is available at \url{https://github.com/AliMohammadiasl/GNNs-for-SLNs}.

\subsection{Empirical Setup} \label{emp_setup}


\begin{table}
    \caption{Metrics for each considered dataset including the number of nodes, edges, duration (weeks), and posts.}
    \centering
    \begin{tabular}{c|c|c|c|c}
    \toprule
        \rowcolor{gray!30}  \textbf{Dataset} & \# \textbf{Nodes} & \# \textbf{Edges} & \textbf{Duration} & \# \textbf{Posts} \\
         \midrule
         \rowcolor{shakeCol} \texttt{VS} & 677 & 4702 & 5 & 7484 \\
         \addlinespace
        \rowcolor{mlCol} \texttt{ML} & 3290 & 30610 & 12 & 25481 \\
        \addlinespace
        \rowcolor{algoCol} \texttt{AL} & 1165 & 6773 & 13 & 16276 \\
        \addlinespace
        \rowcolor{compCol} \texttt{CP} & 900 & 3418 & 8 & 8255 \\
        \bottomrule
    \end{tabular}
    \label{tab:dataset_stats}
\end{table}

For our empirical analysis, we employ four SLNs collected from four distinct Massive Open Online Course (MOOC) datasets:  Virtual Shakespeare (\texttt{VS}), Machine Learning (\texttt{ML}), Algorithms (\texttt{AL}), and English Composition (\texttt{CP}). The SLN data we used was introduced in \cite{InfocomeDataset} and captures structured interactions that occurred between students on discussion forums. Specifically, if student $v_{i}$ and $v_{j}$ interact on a discussion forum by posting an answer or a follow-up question to an initial post, we consider $A_{ij} = 1$. Our considered datasets intentionally span both STEM and humanities courses so that the efficacy of our approach can be evaluated across various disciplines. Tab.~\ref{tab:dataset_stats} summarizes these datasets.

Our GraphSAGE model consists of an $F = 16$-dimensional embedding for each node at the input followed by two convolutional layers. We use Rectified Linear Unit (ReLU) as the activation function following both hidden layers. For training, we perform full-graph updates for a total of 1500 epochs, optimizing the model with the Adam optimizer at a learning rate of $5e-4$ using the above-described weighted binary-cross entropy loss function. 

To ensure robustness (in terms of the time that the model is used during the term/semester), we evaluate our models in multiple temporal snapshots of the SLNs. 
We define progress as the percentage of the term that has elapsed, with 0\% representing the start and 100\% the end. We consider 25\%, 50\%, 75\%, and 90\% progress, and we perform temporal $k$-fold cross-validation (with $k=10$) and report the average AUC across all partitions for statistical validation. 

\subsection{Temporal Graph Analysis} \label{tmp}


\begin{table}[t]
  \centering
  \caption{AUC performance of each considered model across four stages of course progress (25\%, 50\%, 75\%, 90\%). Temporal trends can be observed across progress stages of each course, while spatial effects are seen by comparing `GNN iso' and `GNN all' at each progress stage.}
  \label{tab:auc_all}
  \begin{tabularx}{\columnwidth}{%
      >{\centering\arraybackslash}p{1cm}  
      >{\centering\arraybackslash}p{1.0cm}  
      >{\centering\arraybackslash}p{0.75cm}  
      >{\centering\arraybackslash}p{0.75cm}  
      >{\centering\arraybackslash}p{0.75cm}  
      >{\centering\arraybackslash}p{0.75cm}  
      >{\centering\arraybackslash}p{0.75cm}  
    }
    \toprule
    \rowcolor{gray!30}
    \textbf{Dataset} & \textbf{Progress} & \textbf{GNN all} & \textbf{GNN iso} & \textbf{GCN \cite{Kipf_GCN_2017}} & \textbf{GAT \cite{Velickovic_GAT_2018}} & \textbf{CNN \cite{OG_Paper}} \\
    \midrule
    \rowcolor{shakeCol} \texttt{VS} & 25\% & \textbf{0.88} & 0.86 & 0.82 & 0.87 & 0.64 \\
    \rowcolor{shakeCol} \texttt{VS} & 50\% & \textbf{0.92} & 0.89 & 0.85 & 0.89 & 0.62  \\
    \rowcolor{shakeCol} \texttt{VS} & 75\% & \textbf{0.94} & 0.88 & 0.86 & 0.90 & 0.60  \\
    \rowcolor{shakeCol} \texttt{VS} & 90\% & \textbf{0.94} & 0.88 & 0.85 & 0.89 & 0.67 \\
    \addlinespace

    \rowcolor{mlCol} \texttt{ML} & 25\% & \textbf{0.85} & 0.84 & 0.76 & 0.82 & 0.72  \\
    \rowcolor{mlCol} \texttt{ML} & 50\% & \textbf{0.86} & 0.85 & 0.70 & 0.83 & 0.83 \\
    \rowcolor{mlCol} \texttt{ML} & 75\% & 0.86 & 0.84 & 0.69 & 0.83 & \textbf{0.90}  \\
    \rowcolor{mlCol} \texttt{ML} & 90\% & \textbf{0.88} & 0.83 & 0.87 & 0.84 & 0.81  \\
    \addlinespace

    \rowcolor{algoCol} \texttt{AL} & 25\% & \textbf{0.92} & 0.92 & 0.83 & 0.89 & 0.82 \\
    \rowcolor{algoCol} \texttt{AL} & 50\% & \textbf{0.94} & 0.93 & 0.86 & 0.90 & 0.91  \\
    \rowcolor{algoCol} \texttt{AL} & 75\% & \textbf{0.94} & 0.93 & 0.86 & 0.92 & 0.93 \\
    \rowcolor{algoCol} \texttt{AL} & 90\% & \textbf{0.95} & 0.94 & 0.86 & 0.92 & 0.93  \\
    \addlinespace

    \rowcolor{compCol} \texttt{CP} & 25\% & \textbf{0.86} & 0.84 & 0.82 & 0.83 & 0.66  \\
    \rowcolor{compCol} \texttt{CP} & 50\% & \textbf{0.90} & 0.84 & 0.85 & 0.85 & 0.71  \\
    \rowcolor{compCol} \texttt{CP} & 75\% & \textbf{0.89} & 0.85 & 0.83 & 0.85 & 0.80  \\
    \rowcolor{compCol} \texttt{CP} & 90\% & \textbf{0.91} & 0.88 & 0.85 & 0.85 & 0.60  \\

    \bottomrule
  \end{tabularx}
\end{table}

We begin by examining how link prediction in SLNs evolve temporally throughout the duration of a single course (addressing \textbf{Hypothesis 1} from Sec.~\ref{sec:Introduction}). 
Inspecting Tab.~\ref{tab:auc_all} and focusing on isolated GNN training (i.e., `GNN iso'), which represents GNNs trained on a particular course in isolation, we observe consistent AUC improvements as more temporal data become available in each course. As expected, the AUC improves steadily as the course progresses, and the model benefits from richer structural information in later stages, particularly for smaller courses (e.g., \texttt{VS}, \texttt{AL}, and \texttt{CP}), which suffer from link formation data early on, than larger ones (\texttt{ML}). 
This indicates that the model becomes increasingly capable of distinguishing between likely and unlikely links between students as it observes more interactions. 
These trends indicate that temporal progress has a measurable impact on model performance, affirming \textbf{Hypothesis 1} from Sec.~\ref{sec:Introduction}. Overall, later snapshots provide more robustness for learning, while early-course graphs pose prediction challenges due to the limited structural context. 

\subsection{Spatial Graph Analysis} \label{spatial}

We next analyze the spatial evolution of link prediction in SLNs by exploring how combining SLNs across different classrooms affects model performance (addressing \textbf{Hypothesis 2} from Sec.~\ref{sec:Introduction}). 
To evaluate this, in  Tab.~\ref{tab:auc_all}, we compare the performance of the models when trained on individual SLNs versus on combined SLNs (i.e., `GNN iso' and `GNN all'). 
The results
reveal how merging SLNs positively impacts model performance across different stages of the course. For example, the AUC for \texttt{VS} improves from an average of 0.88  to an average of 0.94 when combined with others over the course progression. 
Moreover, our results show that, although combining SLNs generally improves AUC performance, particularly in smaller or sparser datasets like \texttt{VS} and \texttt{CP}, the gains are less pronounced in larger classrooms such as \texttt{AL} and \texttt{ML}, where the SLN topological structure is already rich and adding more data to it may not affect its topological significance in a meaningful way. This suggests that the benefits of spatial aggregation are not solely due to having more data, but rather depend on the compatibility of interaction patterns across classrooms. 
These results support \textbf{Hypothesis 2} from Sec.~\ref{sec:Introduction}, affirming and revealing the importance and timing of spatial aggregation. 

\subsection{Spatiotemporal Graph Analysis} \label{joint}

\begin{table}[t]
  \centering
  \caption{Hypothesis testing (one-sided $t$-test, $\alpha = 0.10$) comparing GNN isolated (\texttt{IS}) vs.\ GNN combined (\texttt{CO}) vs.\ CNN baseline (\texttt{CNN}) at 25\% and 75\% course progress. Each row compares spatial performance, while reading down the table reveals temporal trends across progress stages.}
  \label{tab:hypothesis-multi}
  \begin{tabularx}{\columnwidth}{%
      >{\centering\arraybackslash}p{1.4cm}  
      >{\centering\arraybackslash}p{.75cm}  
      >{\centering\arraybackslash}p{1.0cm}  
      >{\centering\arraybackslash}p{1.0cm}  
      >{\centering\arraybackslash}p{2.4cm}  
    }
    \toprule
    \rowcolor{gray!30}
    \textbf{Models} & \textbf{Dataset} & \textbf{Progress} & \textbf{$p$-value} & \textbf{Decision} \\
    \midrule
    \rowcolor{algoCol}
     \texttt{IS} vs \texttt{CO} & \texttt{AL} & 25\% & $4.18\text{e}{-1}$ & Fail to Reject $H_0$ \\
    \rowcolor{algoCol}
     \texttt{IS} vs \texttt{CO} & \texttt{AL} & 75\% & $8.03\text{e}{-2}$ & Reject $H_0$ \\
    \rowcolor{algoCol}
     \texttt{CNN} vs \texttt{CO} & \texttt{AL} & 25\% & $6.02\text{e}{-3}$ & Reject $H_0$ \\
    \rowcolor{algoCol}
     \texttt{CNN} vs \texttt{CO} & \texttt{AL} & 75\% & $1.25\text{e}{-1}$ & Fail to Reject $H_0$ \\
    \addlinespace

    \rowcolor{compCol}
    \texttt{IS} vs \texttt{CO} & \texttt{CP} & 25\% & $2.90\text{e}{-2}$ & Reject $H_0$ \\
    \rowcolor{compCol}
    \texttt{IS} vs \texttt{CO} & \texttt{CP} & 75\% & $1.42\text{e}{-2}$ & Reject $H_0$ \\
    \rowcolor{compCol}
    \texttt{CNN} vs \texttt{CO} & \texttt{CP} & 25\% & $5.75\text{e}{-8}$ & Reject $H_0$ \\
    \rowcolor{compCol}
    \texttt{CNN} vs \texttt{CO} & \texttt{CP} & 75\% & $1.23\text{e}{-8}$ & Reject $H_0$ \\
    \addlinespace

    \rowcolor{shakeCol}
    \texttt{IS} vs \texttt{CO} & \texttt{VS} & 25\% & $6.86\text{e}{-4}$ & Reject $H_0$ \\
    \rowcolor{shakeCol}
    \texttt{IS} vs \texttt{CO} & \texttt{VS} & 75\% & $8.00\text{e}{-6}$ & Reject $H_0$ \\
    \rowcolor{shakeCol}
    \texttt{CNN} vs \texttt{CO} & \texttt{VS} & 25\% & $8.13\text{e}{-7}$ & Reject $H_0$ \\
    \rowcolor{shakeCol}
    \texttt{CNN} vs \texttt{CO} & \texttt{VS} & 75\% & $3.7\text{e}{-5}$ & Reject $H_0$ \\
    \addlinespace

    \rowcolor{mlCol}
    \texttt{IS} vs \texttt{CO} & \texttt{ML} & 25\% & $3.06\text{e}{-1}$ & Fail to Reject $H_0$ \\
    \rowcolor{mlCol}
    \texttt{IS} vs \texttt{CO} & \texttt{ML} & 75\% & $6.14\text{e}{-2}$ & Reject $H_0$ \\
    \rowcolor{mlCol}
    \texttt{CNN} vs \texttt{CO} & \texttt{ML} & 25\% & $5.77\text{e}{-2}$ & Reject $H_0$ \\
    \rowcolor{mlCol}
    \texttt{CNN} vs \texttt{CO} & \texttt{ML} & 75\% & $2.15\text{e}{-1}$ & Fail to Reject $H_0$ \\
    \addlinespace

    \bottomrule
  \end{tabularx}
\end{table}

In Secs. \ref{tmp} and \ref{spatial}, we saw that (i) SLNs become structurally richer as a course progresses (temporal evolution) and (ii) combining multiple classroom graphs (spatial evolution) generally boosts AUC by exposing the model to a broader distribution of link patterns. However, these two dimensions do not operate independently. In particular, the benefit of combining SLNs (i.e., using spatial features of the graph) appears to depend on when in the course (temporally) the graphs are merged. Thus, in this section, we  address \textbf{Hypothesis 3} from Sec.~\ref{sec:Introduction}, which investigates whether merging SLNs at different stages of course progress leads to measurable gains in model performance.


As shown in Tab. \ref{tab:auc_all}, we see that combining SLNs earlier in the course leads to higher performance improvements compared to merging them at later stages. This is consistent with the intuition that early in the term, when individual SLNs are sparse, combining information across classrooms helps stabilize learning and improves performance. In contrast, in the later stages of a course, each SLN  develops distinct structural characteristics, potentially reflecting subject matter, classroom dynamics, or group behavior and, as a result, attain lower performance gains upon being combined with other SLNs towards the end of the course. 

Further, to highlight the superiority of our methodology, we compare our GNN-based framework to the current state-of-the-art framework for link prediction in SLNs, which consists of manually engineering fixed graph features and using them as input into a convolutional neural network (CNN) for link prediction \cite{OG_Paper}, along with state-of-the-art approaches for link prediction in general OSNs, which consist of a sum pooling graph convolutional network (GCN) \cite{Kipf_GCN_2017} and a Graph Attention Network (GAT) \cite{Velickovic_GAT_2018}. 
As can be seen from Tab. \ref{tab:auc_all}, each baseline struggles to attain consistently high performance across the spatiotemporal topology of the graph. The CNN in particular struggles due to its fixed feature engineering, whereas the GNN is able to  offer high link prediction performance due to its dynamic adaptation to the SLN graph topology (via its node embeddings) and its evolution over time. 
This is highlighted in Tab. \ref{tab:auc_all}, where the GNN trained on a single SLN (GNN iso) generally outperforms the CNN baseline across all datasets and time instances.
Furthermore, the GNN trained on combined SLNs (GNN all) not only outperforms the CNN but also significantly surpasses the isolated GNN as well as the other baselines in many cases, particularly in earlier stages of small courses. This demonstrates the dual benefit of our framework: first, to outperform traditional feature-based methods, and second, to exploit shared link formation patterns across SLNs to enhance predictive performance. 

To further support the above observations, we perform formal hypothesis testing using a one-sided $t$ test ($\alpha = 0.10$) comparing the AUC performance of our GNN model trained in combination (\texttt{CO}) with other courses against both the GNN trained in isolation (\texttt{IS}) on a single course and the CNN baseline (\texttt{CNN}). 
Here, we focus on the CNN baseline to demonstrate the statistical significance of using our framework's learned features via node embeddings in comparison to fixed feature engineering as currently done in state of the art link prediction frameworks \cite{OG_Paper}.
The null hypothesis in each test states that \texttt{CO} does not outperform the baseline, while the alternative asserts that \texttt{CO} achieves a higher AUC. Tab.~\ref{tab:hypothesis-multi} summarizes the results in 25\% and 75\% course progress for the four SLNs. We show results at these two progress points to highlight the superiority of our method in an early and a late stage of the course. Here, we see that the improvement of our framework is statistically significant, particularly for small courses (e.g., \texttt{CP} and \texttt{VS}). Larger courses (e.g., \texttt{AL} and \texttt{ML}), on the other hand, contain sufficient training data that can be captured in varying time instances. 
This is largely due to the bias of the larger courses as their SLNs account for a larger partition of the spatially aggregated graph structure.  
Yet, we still observe statistically significant gains in AUC when comparing \texttt{CO} against both \texttt{CNN} and \texttt{IS}. In particular, the $p$-values are especially low for smaller and sparser datasets such as \texttt{VS} and \texttt{CP}, even at early progress stages. 
This further supports \textbf{Hypothesis~3}: the spatiotemporal consideration of SLNs generally yields stronger link prediction performance compared to the current state of the art, particularly in smaller courses (\texttt{CP} and \texttt{VS}) that suffer from an abundance of links early in the course. 



\section{Conclusion} \label{conslusion}

In this work, we developed a link prediction framework in SLNs through the lens of two critical dimensions: temporal evolution and spatial aggregation. Using multiple SLNs at varying progress stages, we analyzed how SLNs develop over time and how combining multiple SLNs affects model performance. Our results show that model performance improves as the SLNs progress through the course timeline: richer interaction structures at later stages offer stronger signals for link prediction. We also found that merging SLNs from multiple classrooms can enhance model performance as spatial aggregation helped overcome data sparsity in smaller courses with lower numbers of links throughout the course progression. Finally, our analysis of the spatiotemporal aspects revealed that the benefits of merging graphs are most pronounced early in the course, and particularly in smaller courses, which require more structure early on. 

\section{Acknowledgment}
This work was supported in part by the U.S. National Science Foundation (NSF) under Grant No. SaTC-2513164, ECCS-2512911, ECCS-2543754, and DRL-2418658.


%
\bibliographystyle{abbrv}
\bibliography{sigproc}  
%

\end{document}